\documentstyle[eqsecnum,aps,prb]{revtex}

\newcommand{\rf}[1]{(\ref{#1})}
\newcommand{\mb}{\begin{equation}}
\newcommand{\me}{\end{equation}}
\input psfig.tex
\begin{document}
\draft
\title{Classical artificial two-dimensional atoms: the Thomson model}
\author{B. Partoens\cite{bart} and F. M. Peeters\cite{francois}}
\address{Departement Natuurkunde,
Universiteit Antwerpen (UIA),\\ Universiteitsplein 1, B-2610
Antwerpen, Belgium}
\date{\today}
\maketitle
\begin{abstract}
The ring configurations for classical two-dimensional (2D) atoms are
calculated within the Thomson model and compared with the results from
`exact' numerical simulations. 
The influence of the functional form of the 
confinement potential and the repulsive interaction potential between the
particles on the configurations is investigated. We also give exact
results on those eigenmodes of the system whose frequency does not
depend on
the number of particles in the system.
\end{abstract}
\par

\section{Introduction}
Recently, a detailed investigation of the ground state configurations of a
finite two-dimensional (2D) classical system of charged particles, 
confined by a parabolic external potential, was made in
Ref.~\onlinecite{bedanov}. It was found that the particles arranged
themselves on rings. 
A study of the spectral properties of this classical system such as the 
energy spectrum, 
the eigenmodes, and the density of states was made in 
Ref.~\onlinecite{schweigert}. In both papers Monte Carlo simulations were 
used to study these {\it classical atoms}. 
\par
In the present paper we aim to obtain analytic results for these classical 2D atoms by 
using
a model system: the Thomson 
model. Thomson proposed this classical model in 1904 in order to calculate
the structure 
of the atom.~\cite{thomson} 
He was unable to obtain analytic results for the case of a real 3D
atom and therefore
constrained the particles to move in 
a plane. 
Together with the parabolic confinement potential this is now a model
for the 
classical 2D atoms which were studied in Refs.~\onlinecite{bedanov}
and~\onlinecite{schweigert}.
A crucial {\it ansatz} in the Thomson model is that the particles arrange
themselves in a ring structure. 
Next, Thomson puts as many particles as possible on a single ring, as 
allowed by stability arguments, and the 
rest of them are placed in the centre. By successive applying this
approach he was
able to construct  the 
approximate atomic ring configurations. 
\par
Here we apply the Thomson model and compare the obtained configurations with those
from the 
`exact' numerical simulations of Ref.~\onlinecite{bedanov} and determine when the Thomson
model breaks down. In a second 
step we generalize the Thomson model to: 1) general $r^n$ confinement potentials, and 
2) different functional forms ($1/r^{n'}$ and logarithmic) for the repulsive
inter-particle potential. 
\par
This paper is organized as follows. The calculation to obtain the
structure and
the eigenfrequencies in the Thomson model is outlined in Section~\ref{s2}. We 
compare 
the results for parabolic confinement and Coulomb repulsion with the
previous results from `exact' 
numerical simulations. Section~\ref{s3} is devoted to the extensions of the Thomson
model. 
In Section~\ref{extra} we give some analytical results for the
eigenfrequencies of artificial atoms.
Our conclusions and our results are summarized in Section~\ref{s4}.
 
\section{The Thomson model}
\label{s2}
\subsection{General system}
We study a system of a finite number, $N$, of charged particles
interacting
through a repulsive inter-particle potential and moving in two dimensions 
(2D). A
confinement potential keeps the system together. We focus our attention 
on systems
described by the Hamiltonian
\begin{equation}
H=\sum_{i=1}^N\frac{1}{2}m\omega_0^2\left(\frac{r_i}{\lambda}\right)^n+\frac{e^2}
{\epsilon}\sum_{j>i}^N\frac{1}{|\vec{r}_i-\vec{r}_j|^{n'}},
\label{ham}
\end{equation} 
where $m$ is the mass of the particle, $\omega_0$ the radial confinement frequency,
$e$ the particle charge, $\epsilon$ the dielectric constant of the medium the
particles are moving in, and $\vec{r_i}=(x_i,y_i)$ the
position of the
particle with $r_i\equiv |\vec{r_i}|$. For convenience, we will refer
to our charged
particles as electrons, keeping in mind that they can also be ions with charge $e$
and mass $m$.
We can write the Hamiltonian in a dimensionless form if we express the
coordinates, energy, and force in the following
units
\begin{mathletters}
\begin{equation}
r_0=(e^2/\epsilon)^{1/(n+n')}\alpha^{-1/(n+n')},\\
\end{equation}
\begin{equation}
E_0=(e^2/\epsilon)^{n/(n+n')}\alpha^{n'/(n+n')},\\
\end{equation}
\begin{equation}
F_0=(e^2/\epsilon)^{(n-1)/(n+n')}\alpha^{(n'+1)/(n+n')},
\end{equation}
where $\alpha=\frac{{}_1}{{}^2}m\omega_o^2/\lambda^n$. 
\end{mathletters}
All the
results will be given in reduced form, i.e., in dimensionless units. In such units
the Hamiltonian becomes
\begin{equation}
H=\sum_{i=1}^Nr_i^n+\sum_{j>i}^N\frac{1}{|\vec{r}_i-\vec{r}_j|^{n'}}.
\end{equation}
Thus, the energy is only a function of the number of particles $N$, the
power of the confinement potential $n$ and the power of the interaction
potential $n'$. For $n=2$
and $n'=1$ this system reduces to the one studied in 
Refs.~\onlinecite{bedanov} and~\onlinecite{schweigert}. The numerical
values for the parameters $r_0$ and $E_0$ for some typical experimental
systems 
like electrons in quantum dots,\cite{quantumdots} electron bubbles on a 
liquid Helium surface,\cite{liquidhelium} and ions trapped in Penning 
and Paul traps\cite{paultraps}
were
given in Ref.~\onlinecite{bedanov} for the case of parabolic confinement and 
Coulomb repulsion.
\par
In the Thomson model one obtains the groundstate configurations of
this system by
making the
following
assumption: the $N$ electrons are arranged at equal angular intervals around the
circumference of a circle of radius $a$. Then one investigates the
stability of this 
configuration. With other words, small displacements of the electrons out
of their
equilibrium have to remain small in time. Solving the equations of
motion in polar coordinates to first order in these displacements, we
obtain an
equation which determines the
eigenfrequencies $\omega$ of the system
\begin{equation}
\left((n+n')\frac{I(n')}{2^{n'+1}a^{n'+2}}S_N+L_k-L_0+(n-2)\Omega^2-\omega^2\right)
\left(N_0-N_k-\omega^2\right) 
=\left(M_k-2\Omega\omega\right)^2,
\label{pierke}
\end{equation}
where $\Omega$ is a constant angular velocity with which the whole
system rotates around its centre, $I(n')=n'$, and
\begin{mathletters}
\begin{equation}
S_N=\sum_{j=1}^{N-1}\frac{1}{\sin^{n'}({j\pi}/{N})},
\end{equation}
\begin{equation}
L_k=\frac{I(n')}{(2a)^{n'+2}}\sum_{j=1}^{N-1}
\frac{\cos(2kj\pi/N)}{\sin^{n'+2}(j\pi/N)}
\left(n'\sin^{2}(j\pi/N)+1\right), 
\end{equation}
\begin{equation}
N_k=\frac{I(n')}{(2a)^{n'+2}}
\sum_{j=1}^{N-1}
\frac{\cos(2kj\pi/N)}{\sin^{n'}(j\pi/N)}\left((n'+1)\cot^2
({j\pi}/{N})+1\right),
\label{ganzen}
\end{equation}
\begin{equation}
M_k=n'\frac{I(n')}{(2a)^{n'+2}}
\sum_{j=1}^{N-1}
\sin\frac{2kj\pi}{N}\frac{\cos(j\pi/N)}{\sin^{n'+1}(j\pi/N)}.
\end{equation}
where $k$ is an integer between $0$ and $N-1$. 
\end{mathletters}
The derivation of Eq.~(\ref{pierke}) is straightforward and proceeds along
the
lines given by Thomson\cite{thomson} for the case of $n=2$ and $n'=1$. 
How many frequencies yields this
equation? From the onset we notice that if we replace $k$ by $N-k$ 
in Eq.~(\ref{pierke}), the
values of $\omega$ 
differ only in sign, and thus results in the same frequencies.
Consequently
all the
values of $\omega$ can be obtained by taking only $k=0,1,\ldots,(N-1)/2$
if $N$
is odd,
or $k=0,1,\ldots,N/2$ if $N$ is even. Thus, if $N$ is odd there are
$(N+1)/2$ equations of the type~(\ref{pierke}). For $k=0$, $M_{k=0}=0$,
and Eq.~(\ref{pierke}) reduces to a quadratic equation,  
which implies that the number of
roots of
these $(N+1)/2$ equations equals $4\times(N+1)/2-2=2N$,  
i.e. the number of degrees of freedom of the system, as should be.
For $N$ even, there
are $N/2+1$ equations. But $M_k=0$ for $k=0$ and $k=N/2$, and thus two of
these reduce to quadratics. Consequently, the number of roots is
$4\times\left(N/2+1\right)-4=2N$. Thus in both cases the number of roots is
equal to $2N$, the number of degrees of freedom of the electrons in the plane
of motion and therefore all eigenfrequencies are obtained from
Eq.~(\ref{pierke}).
\subsection{Parabolic confinement and Coulomb repulsion}
Let's consider the case of parabolic confinement ($n=2$) and Coulomb repulsion
between the electrons ($n'=1$). Eq.~(\ref{pierke}) for the
eigenfrequencies becomes
\begin{equation}
\left(\frac{3}{4a^3}S_N+L_k-L_0-\omega^2\right)\left(N_0-N_k-\omega^2\right)
=\left(M_k-2\Omega\omega\right)^2.
\end{equation}
If we take $\lambda=1$ in Eq.~(\ref{ham}) the frequency is expressed in the unit
$\omega'=\omega_0/\sqrt{2}$.
\par
\begin{figure}
\centerline{\psfig{figure=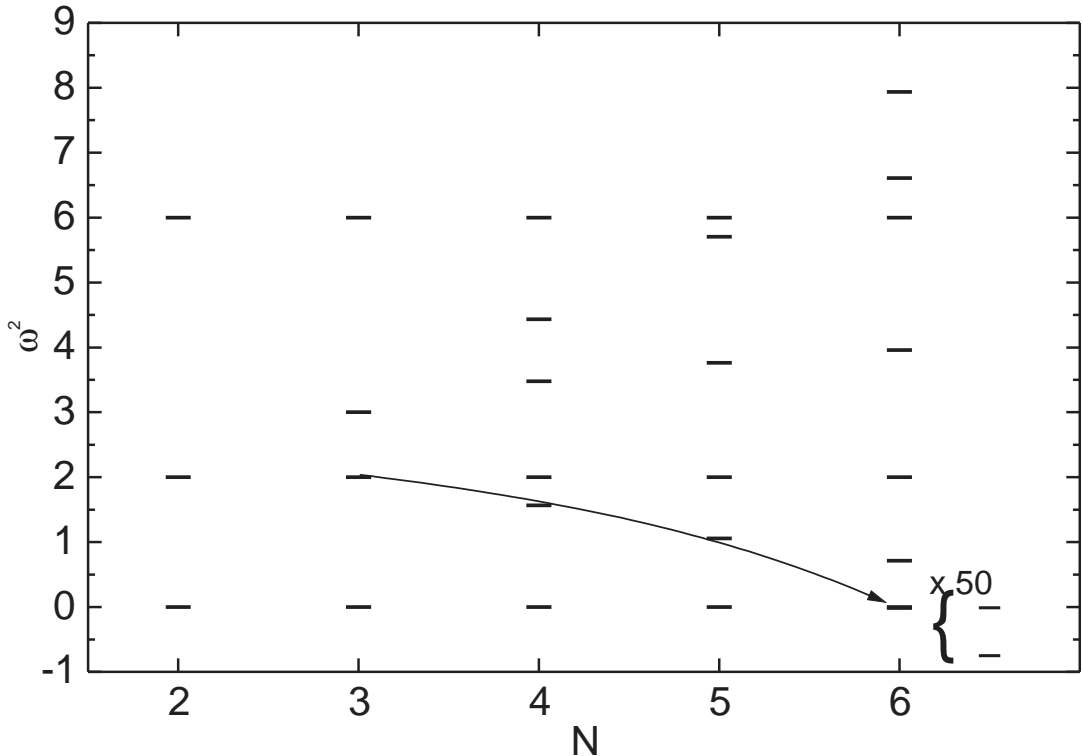,height=10cm}}
\caption{The eigenfrequencies squared for $N=2,3,4,5,6$ for $\Omega=0$.}
\label{softfon}
\end{figure}
\begin{figure}
\centerline{\psfig{figure=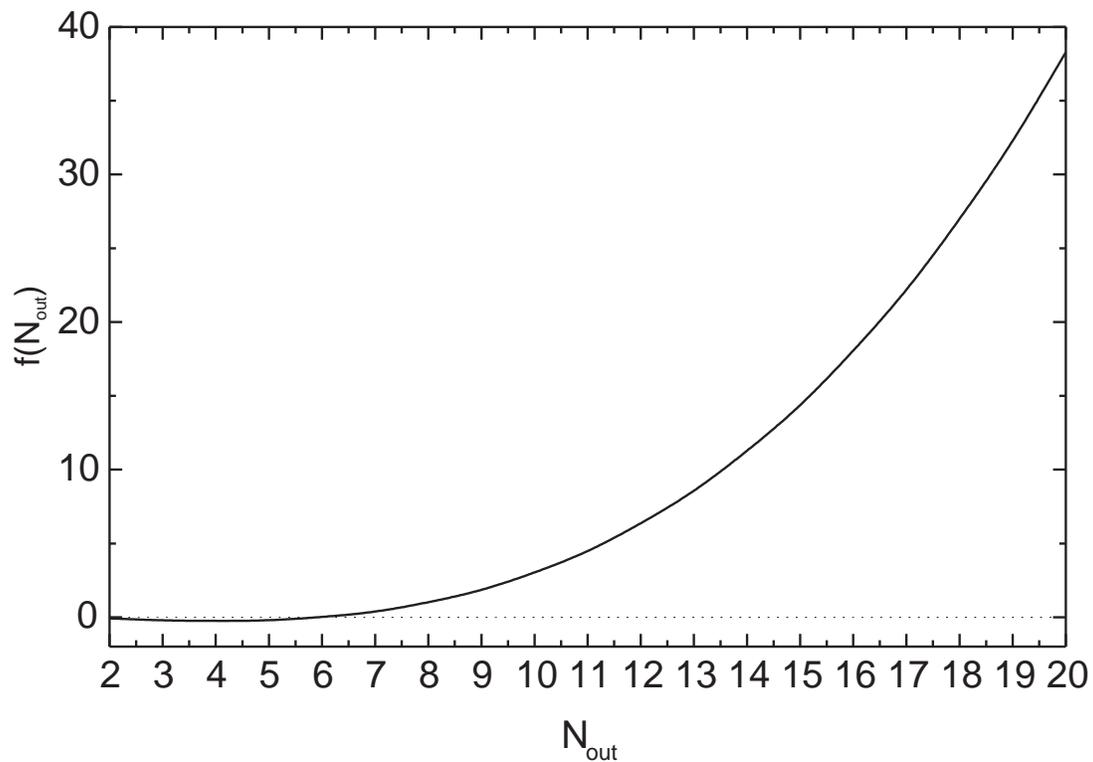,height=10cm}}
\caption{The number of electrons $f(N_{out})$ needed in the centre
of the system in order to stabilize a ring of
$N_{out}$
electrons.}
\label{spline}
\end{figure}

Figure~\ref{softfon} shows the eigenfrequencies squared for $N=1,2,3,4,5$ and $6$ for
zero angular velocity, i.e. $\Omega=0$. Notice that the lowest non zero
eigenfrequency 
decreases and in fact for
$N=6$ we find that $\omega^2<0$ indicating that the single ring configuration 
is no longer stable.
However, it is possible to stabilize the system
by placing electrons in the centre. 
If we put $p$ electrons in the centre of the ring, Eq.~(\ref{pierke})
is modified into
\begin{eqnarray}
\left((n+n')\frac{I(n')}{2^{n'+1}a^{n'+2}}S_N+L_k-L_0+
\frac{(n'+2)I(n')p}{a^{n'+2}}
+(n-2)\Omega^2-\omega^2\right)
\left(N_0-N_k-\omega^2\right) \nonumber\\
=\left(M_k-2\Omega\omega\right)^2.
\label{pierke2}
\end{eqnarray}
From this equation we obtain the minimum value for $p$ which is needed in
order to make a
ring of $N_{out}$ electrons stable. For parabolic confinement and Coulomb repulsion
between the particles we find the condition
\begin{equation}
p>f(k,N_{out})\equiv 
\frac{a^3}{3}\left(L_0-L_k\right)-\frac{S_{N_{out}}}{4}+\frac{a^3}{3}
\frac{M_k^2}{N_0-N_k},
\label{hola}
\end{equation}
which must be satisfied for every $k$. We introduce
$f(N_{out})=
%\max_{0\leq k<N}f(k,N_{out})$
\max_{k\in[0,N)}$
and $p$ equals the integer which is just larger than $f(N_{out})$. 
Figure~\ref{spline} shows the function $f(N_{out})$. 
%The interpolation for
%non-integer $N_{out}$ is done by natural spline functions. 
For $p$ larger than
one, the $p$ inner electrons in principle cannot be
in the same point, i.e. in the centre. 
They will repel each other until they balance the
confinement potential. The Thomson model assumes, as is approximately the
case, that the $p$ electrons around the centre exert the
same force as resulting from $p$
charges placed at the centre. As an example we consider the
case of $N_{out}=12$. For an outer ring of $12$ electrons, 
Eq.~(\ref{hola}) requires that $7$ electrons
are inside (this result can be read off from Fig.~\ref{spline}). But $7$
electrons cannot form a single ring, but will arrange 
themselves as a ring of $6$ with one at the centre. Thus the system of $19$
electrons will consist of an outer ring of $12$, an inner ring of $6$ and one
electron at the centre. 
\par
Following the above procedure we can
construct a table of Mendeljev and compare it with the results
of the `exact' numerical
simulations.~\cite{bedanov} This is done by finding 
the distribution of the electrons when they are
arranged in what we may consider to be the simplest way, i.e. when the number
of rings is a minimum. The number of electrons in the outer ring $N_1$ will
then be determined by the equation
\begin{equation}
N-N_1=f(N_1).
\end{equation}
The value of $N_1$ as obtained from this equation is not an integer 
and consequently  we have to 
take
the integral part of this value. To obtain $N_2$, the number of electrons in the
second ring, we solve
\begin{equation}
N-N_1-N_2=f(N_2).
\end{equation}
We continue in this way until there remain less than $6$ electrons. This procedure
results into 
Table~\ref{table:eenopr}. This table of Mendeljev for the Thomson model
is compared with the
`exact' table of Ref.~\onlinecite{bedanov}. The configurations  
for which the Thomson
model gives the wrong results are printed in italics. 
Notice that this model is capable to predict most of the configurations
correctly. For systems with many electrons ($N>35$) the Thomson
model starts to fail. However even in this case the number of rings is
still predicted correctly until
$N<50$. The reason for this difference is that in the Thomson model the
configurations are found by using a stability
argument, while in Ref.~\onlinecite{bedanov} they were found by a minimalization of
the energy using Monte Carlo simulations. Here we will not compare 
the energy as obtained from the Thomson model with the result given in 
Ref.~\onlinecite{bedanov}. We found that the energy 
in the Thomson model for $N>6$ deviates strongly from the `exact'
results which is due to the assumption that all inner electrons
are placed at the centre. 

\section{Extensions of the Thomson model}
\label{s3}
\subsection{Effect of the confinement potential}
It turned out that for parabolic confinement and Coulomb repulsion
the maximum number of particles on the inner ring is $5$. 
Now we want to investigate how the
confinement potential influences the possible configurations. For
simplicity, we take
$n'=1$ for the inter-particle repulsion, i.e. Coulomb repulsion. 
\par
\begin{figure}
\centerline{\psfig{figure=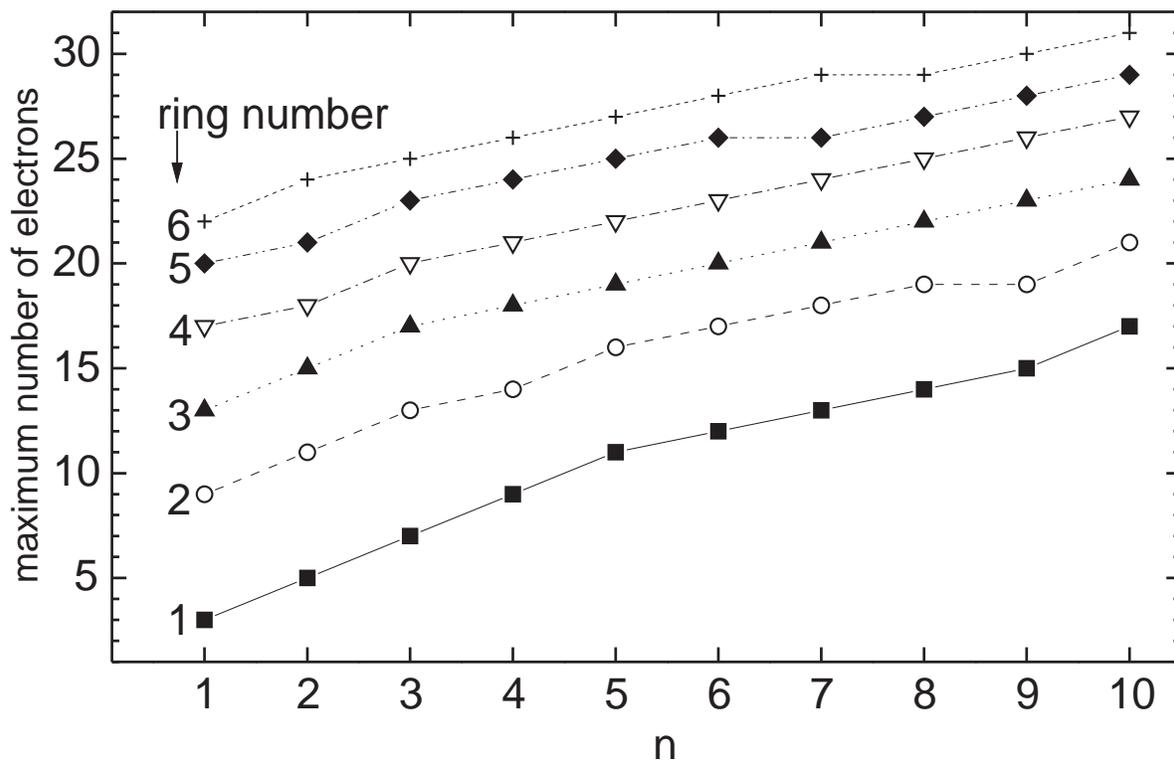,height=10cm}}
\caption{The maximum number of electrons on the inner, second, ..., sixth
ring as function of the power of the confinement potential ($n$) 
for the case of Coulomb repulsion between the electrons.}
\label{conf}
\end{figure}

From Eq.~(\ref{pierke2}) we obtain the minimal value for $p$
which stabilizes a
ring of $N_{out}$ electrons
\begin{equation}
p>\frac{a^3}{3}\left(L_0-l_k\right)-\frac{n+1}{12}S_N+
\frac{a^3}{3}(2-n)\Omega^2+
\frac{a^3}{3}
\frac{M_k^2}{N_0-N_k},
\end{equation}
and this inequality must hold for every $k$. Notice that only for $n=2$,
i.e. parabolic
confinement, this stability condition does not depend on the angular
velocity $\Omega$. For a linear confinement potential ($n=1$) each
configuration will become unstable for sufficient large $\Omega$, while
for
$n>2$ every configuration can be stabilized if the system rotates fast enough. 
To get an idea of the influence of the confinement
potential on the configurations we limit ourselves to  the case
$\Omega=0$. Following 
the same procedure as used
in previous section we find the table of Mendeljev (Table~\ref{knoere}).
The results presented in Table~\ref{knoere} are a good first approximation to the
`exact' results for the considered $N$-values.
Figure~\ref{conf} shows the maximum number of electrons on the
inner, second, ..., sixth ring as function of $n$.
Notice that for a linear confinement potential the inner ring
can support a maximum of $3$ electrons which is much smaller than
in the
case of a parabolic confinement. For confinement potentials with
$n>2$ it's just the opposite, more electrons can be fitted on the rings
without destabilizing them.
\par
In the limit of $n\rightarrow\infty$ the confinement potential becomes a
hard wall potential, i.e. $V(r)=0\;\; (r<\lambda), \infty
\;\;(r>\lambda)$. We
found numerically that the maximum number of particles on each of the
rings keeps increasing with $n$. This clearly signals the breakdown
of the Thomson model in this limit because from `exact' numerical
simulations (see
Ref.\onlinecite{bedanov}) with a hard wall circular potential it is known
that, with increasing number of
electrons, inner rings are formed. We can understand this as follows. In
the Thomson model the groundstate configurations are found using a
stability argument, so only inner rings are formed to stabilize the outer
electrons. For a hard wall potential, there is no confining force for
$r<\lambda$, thus no inner electrons are needed. All the electrons form
one ring at $r=\lambda$, but this is not the configuration with the
minimum value of the energy.
\par
For completeness, we mention that recently Farias and Peeters~\cite{farias}
studied a system of $4$ electrons in a Coulomb type of confinement and found 
different configurations depending on the strength of the Coulomb
confinement potential. This together with the present results indicate that 
the number of electrons on each ring are not universal but can be strongly
influenced by the type and strength of the confinement potential.
\subsection{Effect of the inter-particle interaction potential}
Now we investigate the effect of the functional form of the 
interaction potential between the
electrons on the groundstate configurations. For simplicity, we consider 
here a
parabolic confinement potential ($n=2$).
\par
First we can extend the inter-particle potential to a logarithmic
interaction. To obtain Eqs.~\rf{pierke} and~\rf{pierke2} we used
$1/r^{n'}$
as the inter-particle potential. These equations are obtained from the law of Newton 
which contains the force. Because the derivative of $\ln r$ equals
$1/r$ these equations are therefore also valid for a logarithmic
inter-particle interaction where we have to take
$n'=0$ and $I(0)=1$.
\par
\begin{figure}
\centerline{\psfig{figure=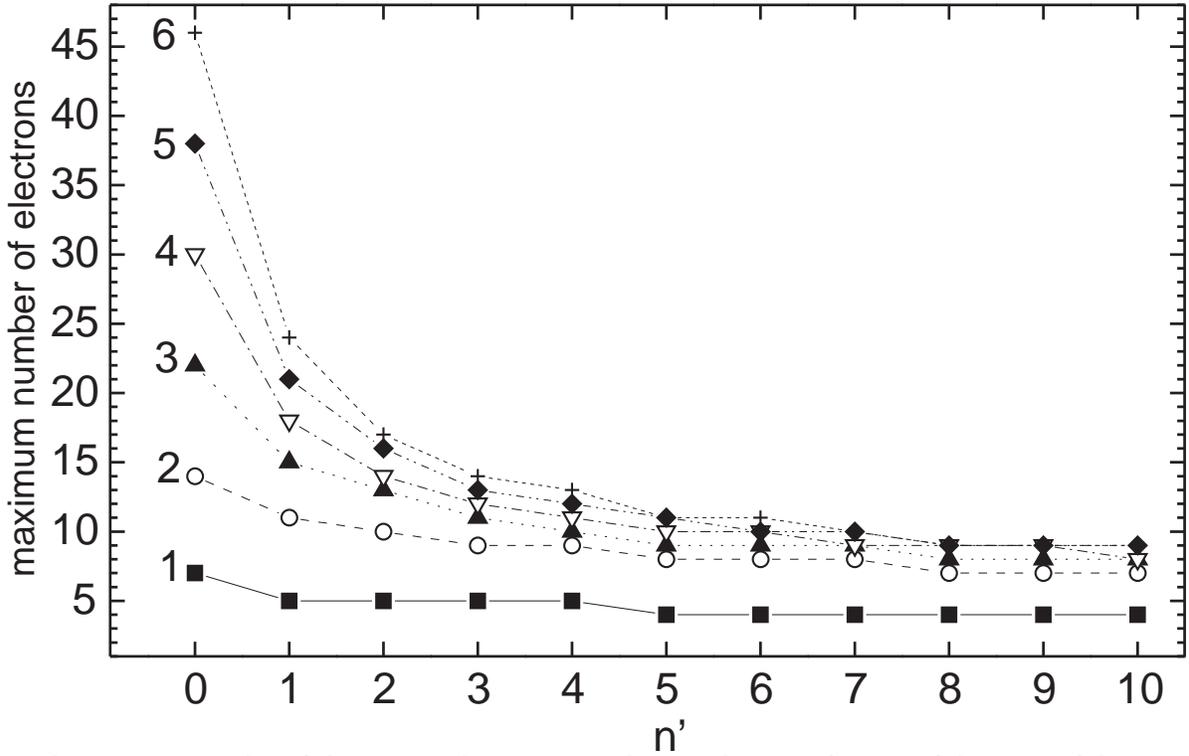,height=10cm}}
\caption{The maximum number of electrons on the inner, second, ..., sixth
ring as function of the power of the inter-particle interaction ($n'$) for
the case of a parabolic confinement potential. $n'=0$ corresponds
to a logarithmic interaction.}
\label{int}
\end{figure}

In Table~\ref{lamp}, we give the table of Mendeljev as obtained 
from the generalized Thomson model using a
logarithmic and a $1/r^2$ interaction, and compare the results with the one for
a Coulomb interaction (Table~\ref{lamp}). 
For the $1/r^2$ interaction we 
find that with increasing number of electrons
more electrons have to be placed
at the centre than for the case of Coulomb interaction in order to stabilize the
outer ring. 
Therefore the resemblance with the `exact' 
table of Mendeljev will not be as good 
as for the case of Coulomb interaction. Indeed, the effect of the
assumption that all inner 
electrons are placed at the centre will be larger. For the 
logarithmic interaction the opposite is found. 
Fewer electrons have to be placed at the centre and in this case the Thomson
model should work very well. Figure~\ref{int} shows the maximum number of
electrons on the inner, second, ..., sixth ring as function of $n'$.
Notice that the maximum occupation number of the different rings decreases
with increasing $n'$. 
\par
For $n'\rightarrow\infty$ we found
numerically $N_1=4$ and for
all the outer rings $N=6$. Also in this limit we do not expect that the
Thomson model gives the correct results, because the
inter-particle
interaction becomes extremely short range, i.e. they are delta function
like. Classically
the electrons can sit very close to each other and thus will occupy the
$r\approx0$ region where the confinement potential is zero. As a
consequence, such a system is similar to the case of an infinite 2D Wigner
lattice where the particles form a hexagonal lattice, i.e. each electron
has $6$ neighbours, and no ring structure is expected. Similar results
were recently obtained~\cite{studart} for a screened Coulomb interaction
$e^{-\kappa r}/r$ using a molecular dynamics simulation approach. With
increased screening, i.e. larger $\kappa$, structural transitions were
found in which the ring configurations changed abruptly and in the limit
of very large $\kappa$ a 2D Wigner type lattice was formed in the centre
of the classical 2D atom.

\section{Eigenfrequencies independent of the number of
electrons} \label{extra}
Figure~\ref{softfon} suggests that in the case of parabolic confinement
and
Coulomb repulsion there are three eigenfrequencies independent of the
number of electrons (see also Ref.~\onlinecite{schweigert}). The existence
and
value of these eigenfrequencies can
be obtained analytically.
1) For any axial symmetric system the system as a whole can rotate,
which leads to an eigenfrequency $\omega=0$.
2) The Hamilton equation of motion yields $\dot{v}_{xi}=-n x_i
\left(x_i^2+y_i^2\right)^{n/2-1}+
n'\sum_{j_{j\neq i}}(x_i-x_j)/{r_{ij}^{n'+2}}$. Now consider the
centre of mass $\vec{R}=\sum_i\vec{r}_i$ which satisfies the differential
equation 
\begin{equation}
\frac{d^2R_x}{dt^2}=\sum_i\dot{v}_{xi}=-n\sum_ix_i\left(x_i^2+y_i^2
\right)^{n/2-1},
\label{conny}
\end{equation}
and of course the same for $R_y$. Notice that Eq.~(\ref{conny}) is
independent of $n'$.
Thus only for a parabolic confinement
potential the above equation reduces to $d^2R_x/dt^2=-2R_x$ and a twofold
degenerate vibration of the centre 
of mass with eigenfrequency $\omega=\sqrt{2}$ is obtained. This frequency  
is independent of the
number of electrons and independent of the inter-particle potential which 
is a consequence of the generalized Kohn theorem.\cite{kohn}   
3) For the mean square radius $R^2=\sum_i\left(x_i^2+y_i^2\right)$ we
find 
\begin{equation}
\frac{d^2R^2}{dt^2}=-(n+n')\sum_i\left(x_i^2+y_i^2\right)^{n/2} +n'H+2T,
\end{equation}
with $T=\sum_i\left(\dot{x}_i^2+\dot{y}_i^2\right)$ the total kinetic
energy. For parabolic confinement, i.e. $n=2$, there is a breathing mode
with
frequency $\omega=\sqrt{2+n'}$ which is independent of the number of
electrons. The
existence of the breathing mode does not depend on the functional form of
the inter-particle potential, but its value does.

\section{Conclusions and summary}
\label{s4}
The Thomson model, a model for classical 2D atoms, was investigated and the
groundstate configurations were obtained.
The results are valid for repelling particles with arbitrary mass 
and charge. First, we deduced within the Thomson
model a table of
Mendeljev for parabolic confinement and Coulomb repulsion between the
electrons and compared it with the results from `exact' Monte Carlo
simulations.~\cite{bedanov} The Thomson model correctly predicts the
configurations for $N<36$.
This table was constructed using
stability arguments: the eigenfrequencies of the 2D atom are calculated
and determined when one of them becomes imaginary in order to find the
maximum number of electrons on the rings.    
\par
Knowing that the Thomson model is a rather good model to predict the
configurations of classical 2D atoms we investigated the effect of a
$r^n$-confinement potential. Only for parabolic confinement the
configurations do not depend on the angular velocity for rotation of the
total system. For a linear
confinement potential the configuration becomes unstable if the system
rotates too fast, while for $n>2$ every configuration may be stable if the
system rotates fast enough. We also found that the maximum number of
allowed electrons on each ring is an increasing function of $n$,
 i.e. it increases with the steepness of the confinement potential. 
\par
We extended the Thomson model to a $1/r^{n'}$-type and to logarithmic
interaction between
the electrons and found the groundstate configurations. Here we found that with
increasing $n'$, i.e. for shorter range of inter-particle interaction,
fewer electrons
can be put on a ring. 
\par
We showed that there are three
eigenfrequencies which are independent of the number of electrons in the
case of parabolic confinement. The zero frequency mode which is a
consequence of the axial symmetry of the system. We found that only for a
parabolic
confinement potential the frequencies of the vibration of the centre of
mass and the breathing mode are independent of the number of electrons. 
The value of the latter does depend on the functional form of the
inter-particle potential.

\section{Acknowledgments}
This work is 
supported by the Human Capital and Mobility network Programme No. ERBCHRXT
930374. B.P. and F.M.P. are supported by the Flemish Science Foundation
(FWOVlaanderen).

\begin{table}
\caption{Table of Mendeljev for classical 2D atoms. The results as obtained from 
the Thomson model are compared with the `exact' 
results
from the Monte Carlo simulations of Ref.~\protect\onlinecite{bedanov}.}
\begin{tabular}{cll|cll} 
N&Thomson&Monte Carlo&N&Thomson&Monte Carlo\\  \tableline
1&1&1&   26&3,9,14&3,9,14\\
2&2&2&             27&4,9,14&4,9,14\\
3&3&3&             28&4,10,14&4,10,14\\
4&4&4&             29&5,10,14&5,10,14\\
5&5&5&             30&5,10,15&5,10,15\\
6&1,5&1,5&                 31&5,11,15&5,11,15\\
7&1,6&1,6&                 32&1,5,11,15&1,5,11,15\\
8&1,7&1,7&                 33&1,6,11,15&1,6,11,15\\
9&2,7&2,7&                 34&1,6,12,15&1,6,12,15\\
10&2,8&2,8&                35&1,6,12,16&1,6,12,16\\
11&\emph{2,9}&3,8&                36&\emph{1,7,12,16}&1,6,12,17\\
12&3,9&3,9&                37&\emph{2,7,12,16}&1,7,12,17\\
13&4,9&4,9&                38&\emph{2,7,13,16}&1,7,13,17\\
14&4,10&4,10&              39&\emph{2,8,13,16}&2,7,13,17\\
15&5,10&5,10&              40&2,8,13,17&2,8,13,17\\
16&\emph{5,11}&1,5,10&                    41&\emph{2,9,13,17}&2,8,14,17\\
17&\emph{1,5,11}&1,6,10&                  42&\emph{3,9,13,17}&3,8,14,17\\
18&1,6,11&1,6,11&                  43&3,9,14,17&3,9,14,17\\
19&1,6,12&1,6,12&                  44&\emph{4,9,14,17}&3,9,14,18\\
20&1,7,12&1,7,12&                  45&\emph{4,9,14,18}&3,9,15,18\\
21&\emph{2,7,12}&1,7,13&                  46&\emph{4,10,14,18}&4,9,15,18\\
22&\emph{2,7,13}&2,8,12&           47&\emph{5,10,14,18}&4,10,15,18\\
23&2,8,13&2,8,13&                  48&\emph{5,10,15,18}&4,10,15,19\\
24&\emph{2,9,13}&3,8,13&           49&\emph{5,11,15,18}&4,10,15,20\\
25&3,9,13&3,9,13&                  50&\emph{1,5,11,15,18}&4,10,16,20\\
\end{tabular}
\label{table:eenopr}
\end{table}

\begin{table}
\caption{Table of  Mendeljev for the Thomson model for various
confinement potentials $r^n$.}
\begin{tabular}{cllll|cllll}
N&n=1&n=2&n=3&n=10&N&n=1&n=2&n=3&n=10\\ \tableline
3&3&3&3&3&14&1,4,9&4,10&4,10&14\\
4&1,3&4&4&4&15&1,4,10&5,10&4,11&15\\
5&1,4&5&5&5&16&1,5,10&5,11&5,11&16\\
6&1,5&1,5&6&6&17&1,5,11&1,5,11&5,12&17\\
7&1,6&1,6&7&7&18&1,6,11&1,6,11&6,12&1,17\\
8&2,6&1,7&1,7&8&19&2,6,11&1,6,12&7,12&2,17\\
9&2,7&2,7&1,8&9&20&2,7,11&1,7,12&7,13&3,17\\
10&2,8&2,8&1,9&10&21&2,7,12&2,7,12&1,7,13&4,17\\
11&3,8&2,9&2,9&11&22&2,8,12&2,7,13&1,8,13&4,18\\
12&3,9&3,9&2,10&12&23&3,8,12&2,8,13&1,9,13&5,18\\
13&1,3,9&4,9&3,10&13&24&3,8,13&2,9,13&1,9,14&6,18\\ 
\end{tabular}
\label{knoere}
\end{table}

\begin{table}
\caption{The groundstate configurations for logarithmic, Coulomb and
$1/r^2$ interaction between 
the particles, as obtained within the Thomson model, as a function of the
total number of particles $N$.}
\begin{tabular}{clll|llll}
N&$-\ln r$&$1/r$&$1/r^2$&N&$-\ln r$&$1/r$&$1/r^2$\\
\tableline
2&2&2&2&           27&2,9,16&4,9,14&5,10,12\\
3&3&3&3&           28&2,9,17&4,10,14&5,10,13\\
4&4&4&4&           29&2,10,17&5,10,14&1,5,10,13\\
5&5&5&5&           30&2,10,18&5,10,15&1,6,10,13\\
6&6&1,5&1,5&           31&3,10,18&5,11,15&1,6,11,13\\
7&1,6&1,6&1,6&         32&3,11,18&1,5,11,15&1,7,11,13\\
8&1,7&1,7&1,7&         33&3,11,19&1,6,11,15&2,7,11,13\\
9&1,8&2,7&2,7&         34&4,11,19&1,6,12,15&2,8,11,13\\
10&2,8&2,8&2,8&            35&4,12,19&1,6,12,16&2,8,11,14\\
11&2,9&2,9&3,8&            36&4,12,20&1,7,12,16&3,8,11,14\\
12&2,10&3,9&3,9&           37&5,12,20&2,7,12,16&3,8,12,14\\
13&3,10&4,9&4,9&           38&5,13,20&2,7,13,16&3,9,12,14\\
14&3,11&4,10&5,9&          39&5,13,21&2,8,13,16&4,9,12,14\\
15&4,11&5,10&5,10&         40&6,13,21&2,8,13,17&5,9,12,14\\
16&4,12&5,11&1,5,10&           41&6,14,21&2,9,13,17&5,10,12,14\\
17&5,12&1,5,11&1,6,10&         42&6,14,22&3,9,13,17&5,10,13,14\\
18&5,13&1,6,11&1,6,11&         43&1,6,14,22&3,9,14,17&1,5,10,13,14\\
19&6,13&1,6,12&1,7,11&         44&1,7,14,22&4,9,14,17&1,5,10,13,15\\
20&6,14&1,7,12&2,7,11&         45&1,7,15,22&4,9,14,18&1,6,10,13,15\\
21&1,6,14&2,7,12&2,8,11&           46&1,7,15,23&4,10,14,18&1,6,11,13,15\\
22&1,7,14&2,7,13&3,8,11&           47&1,8,15,23&5,10,14,18&1,7,11,13,15\\
23&1,7,15&2,8,13&3,8,12&           48&1,8,16,23&5,10,15,18&2,7,11,13,15\\
24&1,8,15&2,9,13&3,9,12&           49&1,8,16,24&5,11,15,18&2,8,11,13,15\\
25&1,8,16&3,9,13&4,9,12&           50&2,8,16,24&1,5,11,15,18&2,8,11,14,15\\
26&2,8,16&3,9,14&5,9,12&51&2,9,16,24&1,6,11,15,18&3,8,11,14,15\\
\end{tabular}
\label{lamp}
\end{table}
\end{document}